\documentclass[aps,pre,twocolumn,float]{revtex4}
\usepackage{amsmath,bm,epsfig}
\newcommand{\B}[1]{{\bm{#1}}}
\usepackage[dvips]{color}
\begin{document}

\title{Statistics of Plastic Events in Post-Yield Strain-Controlled Amorphous Solids}
\author{Awadhesh K. Dubey$^{1}$, H. George E. Hentschel$^{2}$, Itamar Procaccia$^{1}$ and Murari Singh$^{1}$ }
\affiliation{$^{1}$Department of Chemical Physics, The Weizmann Institute of Science,  Rehovot 76100, Israel\\
$^{2}$Department of Physics,  Emory University - Atlanta, Georgia, USA}
\begin{abstract}
Amorphous solids yield in strain-controlled protocols at a critical value of the strain. For
larger strains the stress and energy display a generic complex serrated signal
with elastic segments punctuated by sharp energy and stress plastic drops having a wide range of magnitudes. Here we
provide a theory of the scaling properties of such serrated signals taking into account the system-size
dependence. We show that the statistics are not
homogeneous - they separate sharply to a regime of `small' and `large' drops, each endowed with its own
scaling properties. A scaling theory is first derived solely by data analysis, showing a somewhat complex picture. But after considering the physical interpretation one discovers that the scaling behavior and the scaling exponents are in fact very simple and universal.
\end{abstract}
\maketitle
\section{Introduction}
Serrated signals are ubiquitous in ``stick-slip" physical systems; examples range from earthquakes \cite{EQ}, through Barkhasuen Noise in magnetic systems \cite{Bark,14HIPS},
to stress and energy as a function of strain in amorphous solids \cite{04VBB,04ML,05DA,06TLB,06ML,09LP,11RTV,06SLG,13KTG,13NSSMM}. The analysis of such signals often tends to seek power-laws
to describe the statistics of the magnitude of serrated events\cite {10Wang,13Wang}. In this paper we stress that a more complete understanding of
the statistics of such phenomena calls for scaling functions which incorporate knowledge of the system-size dependence. The system size dependence reveals crucial information pertaining to the underlying physics responsible for the serrated signals.
We will show that the examination of the statistics of the magnitude of serrated responses together with the system
size dependence may discover important inhomogeneities in the statistics that may escape attention when focusing
on power-laws alone. In typical cases there is more than one physical mechanism contributing to the observed signal, and these
mechanism must be identified \cite{16HPS}. Finally we will show that a full understanding of the statistics of such signals and their scaling properties calls for an examination of the physical processes involved. After such an examination the picture may clarify
considerably and even universal results may be gleaned.
\begin{figure}[ht!]
\includegraphics[width=7cm]{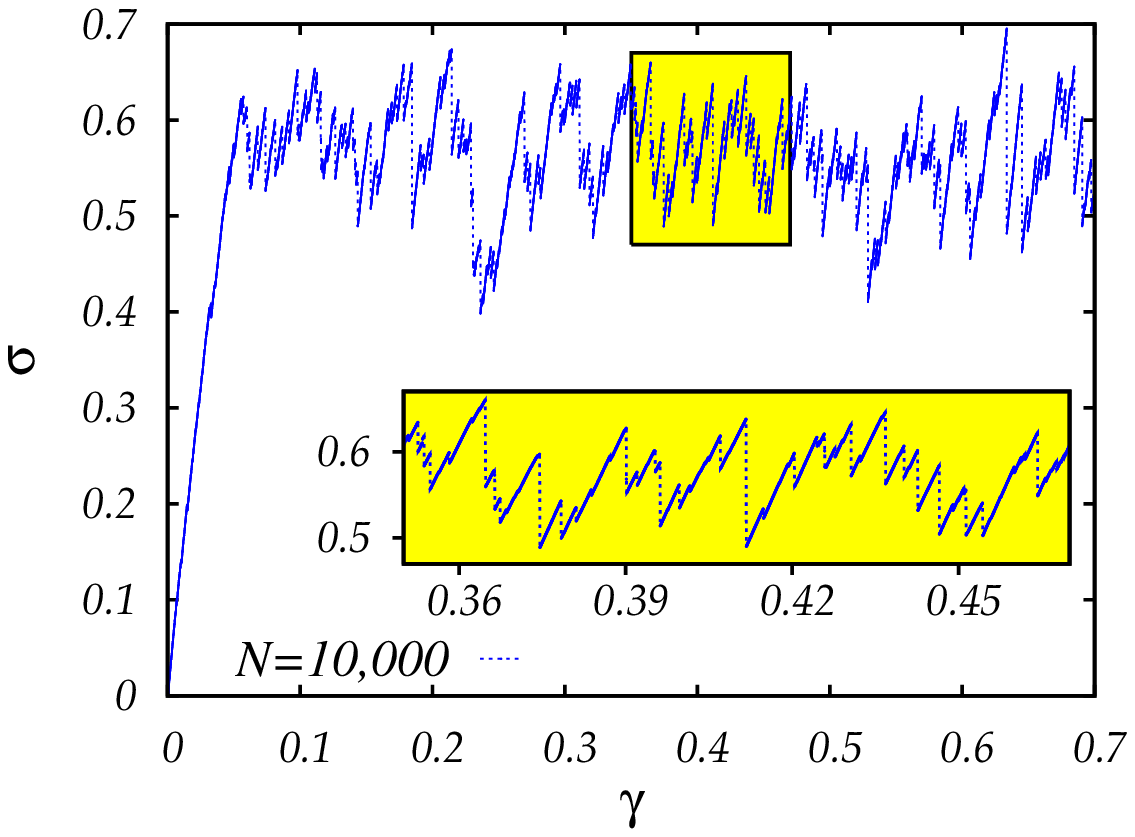}
\includegraphics[width=7cm]{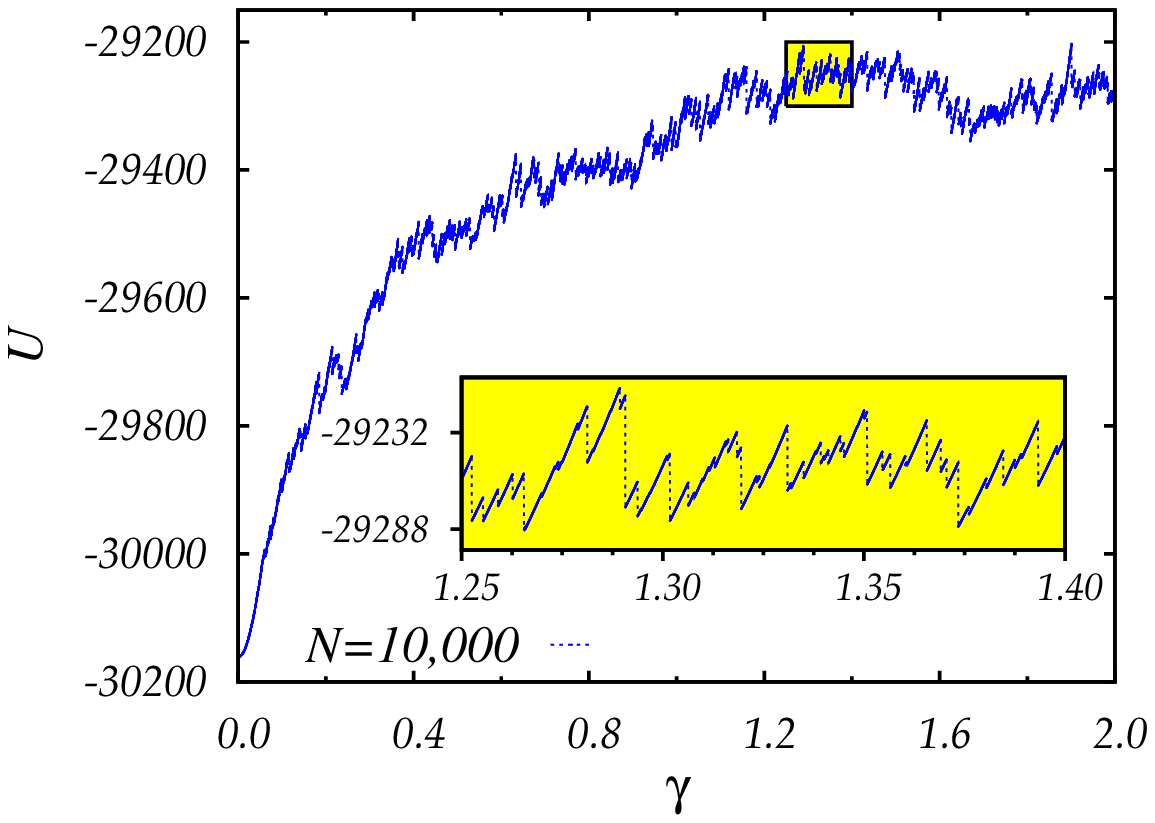}
\caption{Typical stress vs. strain and energy vs. strain curves obtained in a strain-controlled athermal quasistatic (AQS)
shearing protocol. The insets show blown up versions of the shady regions in the main curves.
Note that the stress reaches a steady state much faster than the energy. The average stress is constant
already after 10\% in strain, whereas the energy needs more than 100\% in strain before steady conditions are met. Here the system size is 10,000 particles, and see the text for further simulational details.}
\label{stress-strain}
\end{figure}

The context of our analysis is the mechanical responses of amorphous solids, and the serrated
signals of stress and energy as a function of strain. Thus we will consider the shear stress $\sigma_{xy}$ and the energy $U$ as a function of shear strain $\gamma_{xy}$ in amorphous
solids  \cite{98Ale}. Since we will only concern ourselves with one component of the stress and strain tensors we denote
them below as $\sigma$ and $\gamma$ respectively. Typically the stress vs strain curves and the energy vs strain curves as measured in amorphous solids exhibit two distinct regions. At lower strain values the stress $\sigma$ and the energy $U$ increase on the average upon the increase of strain $\gamma$, but this increase is punctuated by plastic events. A second region, at higher values of the strain,
displays a elasto-plastic steady state in which the stress and energy cannot increase on the average even though the strain keeps increasing. We will present an analysis of the statistics of the sharp drops involving plastic events in this
second region, where the steady-state properties allow us to gather enough data for accurate statistics \cite{10KLPZ}. Typical such shear stress and energy vs. shear strain curves at zero temperature are shown in Fig.~\ref{stress-strain}. The aim of this
 paper is to understand the statistics of the complex looking serrated signal in the post-yield steady state regime \cite{15HJPS}.
 \section{Simulaitons and Data analysis}
 \subsection{System Details}
In this work, we employ a two-dimensional Kob-Andersen \cite{94KA} binary glass-former with a 65:35 ratio
of point particles A and B having equal mass $m$, with interaction given by shifted and smoothed Lennard-Jones (LJ) potentials, $u_{\alpha\beta}(r)$,
\begin{equation}
u_{\alpha\beta}(r) = \begin{cases} u^{LJ}_{\alpha\beta}+A_{\alpha\beta} +B_{\alpha\beta}r+C_{\alpha\beta}r^2, & \mbox{if } r \leq R^{cut}_{\alpha\beta}
                                 \\ 0, & \mbox{if } r > R^{cut}_{\alpha\beta}, \end{cases}
\label{Usmooth}
\end{equation}
 where
\begin{equation}
u^{LJ}_{\alpha\beta} = 4\epsilon_{\alpha\beta}\left[\left(\frac{\sigma_{\alpha\beta}}{r}\right)^{12} - \left(\frac{\sigma_{\alpha\beta}}{r}\right)^6\right].
\label{ULJ}
\end{equation}
The smoothing  of potentials in Eq.(\ref{Usmooth}) is such that they vainsh with
two zero derivatives at distances $R^{cut}_{\alpha\beta} = 2.5\sigma_{\alpha\beta}$.
The parameters for smoothing the LJ potentials in Eq. (\ref{Usmooth}) and
for A and B particle type interactions in Eq.(\ref{ULJ}) are given in the following table
\begin{center}
\begin{tabular}{ |c|c|c|c|c|c| }
\hline
Interaction & $\sigma_{\alpha\beta}$ & $\epsilon_{\alpha\beta}$ & $A_{\alpha\beta}$ & $B_{\alpha\beta}$ & $C_{\alpha\beta}$ \\
 \hline
 AA & 1.00 & 1.0 & 0.4527 & -0.3100 & 0.0542 \\
 BB & 0.88 & 0.5 & 0.2263 & -0.1762 & 0.0350 \\
 AB & 0.80 & 1.5 & 0.6790 & -0.5814 & 0.1271 \\
 \hline
\end{tabular}
\end{center}
The reduced units for mass, length, energy and time have been taken as $m$,
$\sigma_{AA}$, $\epsilon_{AA}$ and $\sigma_{AA}\sqrt{m/\epsilon_{AA}}$ respectively.

\subsection{Preparation of Amorphous Solids}
 To prepare amorphous solid, we start with a configuration generated randomly
 at $\rho=$1.162 and then equilibrate it using molecular dynamics (MD) technique at higher temperature $T=0.4$ for 400,000 MD steps.
 Next, we cool down the system, with cooling rate of  $\dot T=10^{-6}$ in reduced units, to  desired temperature of $T=0.000001$.
 Finally we instantly quench the configuration  obtained at $T=0.000001$ to the nearest inherent minima at $T=0$ using conjugate gradient minimization
 technique. We repeat this process starting from different initial conditions at $T$ = 0.4 to generate the ensemble of 1000 amorphous solids
 at each system size. The different system sizes used for the analysis are $N=$200, 500, 1000, 2000, 4000, and 10,000.

 \subsection{Athermal Quasistatic (AQS) Protocol}
 Once we have the ensemble of amorphous solids at each system size, we strain every of amorphous solid in an athermal quasistatic (AQS) limit, $T\rightarrow 0$
 and $\dot\gamma\rightarrow 0$, to  examine their stress-strain curve and to collect the statistics of post-yield plastic events. In each step of this method,
 the particle positions are subjected to the affine transformation,
\begin{equation}
x_i \rightarrow x_i + y_i \delta \gamma \ ,\quad y_i \rightarrow y_i \ .
\end{equation}
with Lees-Edwards coundary conditions\cite{91ATBook}. The change in the particle positions due to affine transformation put the system out of
mechanical equilibrium due to amorphous nature of system, and we therefore allow the second  step to the particle positions,
a nonaffine transformation $\B r_i \rightarrow \B r_i + \B u_i$ which annuls the forces between the particles, returning back the system
to  mechanical equilibrium. We choose the basic strain increment step to be $\delta \gamma= 5\times10^{-5}$ for all system sizes simulated.

On increasing the external strain, when the system's reversible elastic branches are terminated by mechanical instability, a plastic event is detected. We measure the
corresponing  energy drop $\Delta U=U_{before}-U_{after}$ and stress drop $\Delta \sigma=\sigma_{before}-\sigma_{after}$. On the other hand $\Delta \gamma$ is defined
as the interval between two successive plastic events.
In order to increase the precision in determining the locations and values of stress and energy drops in the elastoplastic steady state
and to ensure that we do not overshoot and miss the next plastic event, we stop the simulation after a drop is detected, backtrack
to the configuration prior to the drop, and we use much smaller strain increment of $\delta \gamma= 1\times10^{-7}$ till the drop is detected.
Now to collect the statistics of plastic drops in the steady state, we keep on straining the sytem till the system reaches stationarity.
In the present simulation, we observe the stationarity in the system after 100\% straining. For the present study, we collect
the data corresponding to plastic drops from  200\% to 300\% of straining.

 \subsection{Data analysis}
 The steady state branch itself is independent of the preparation protocol; we will collect data from the region where the memory of the initial state is lost. Note that the steady state in stress appears to set in much before the steady state in energy, which keeps
increasing beyond the `yield' point at which the stress achieves a steady state. It is important to wait for true
steady state to collect precise stationary statistics. We judge the stability of the steady state by computing the mean
value of the stress and energy, denoting below as $\sigma_\infty$ and $U_\infty$, making sure that they reach stationary values.

\begin{figure}
\includegraphics[width=7cm]{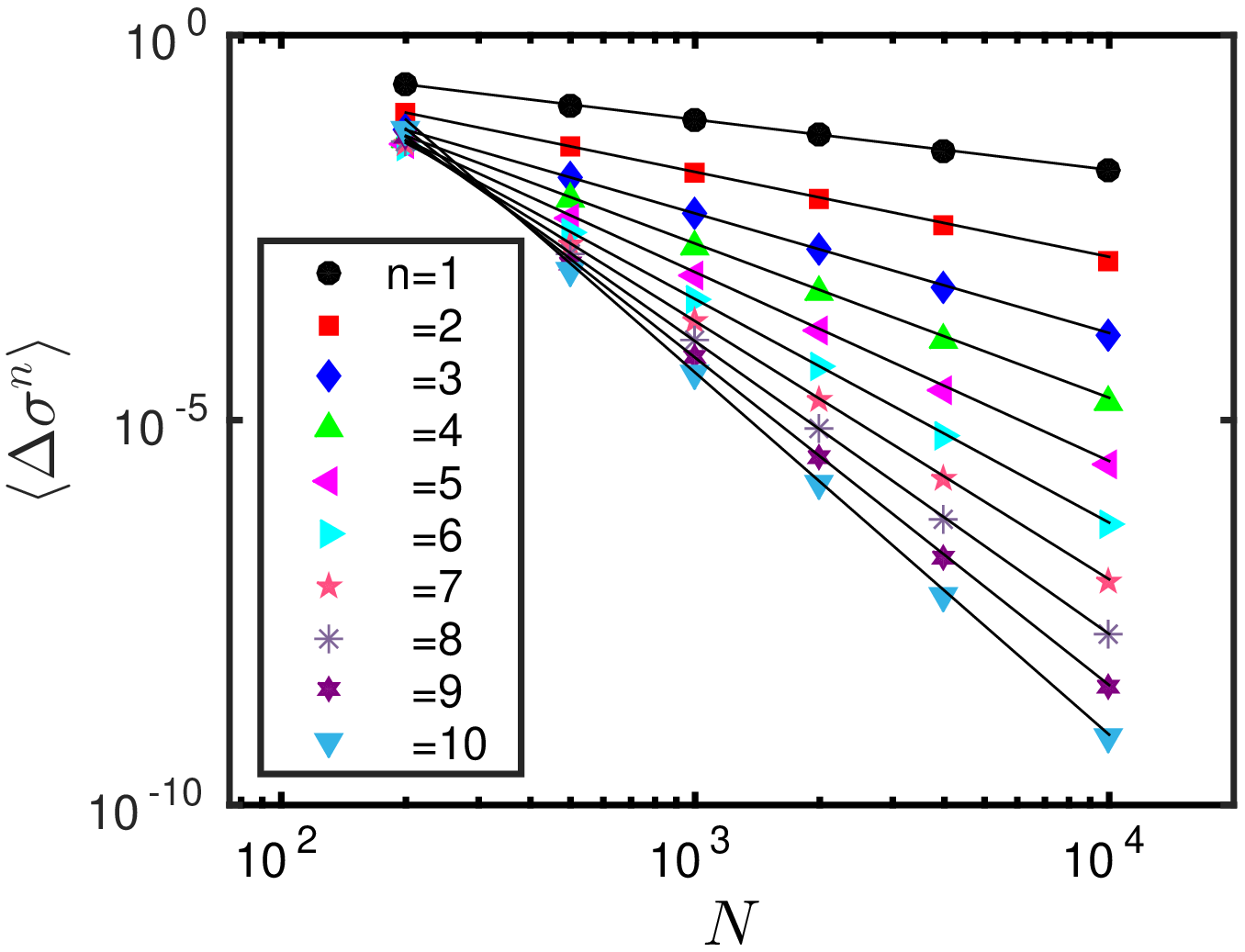}
\includegraphics[width=7cm]{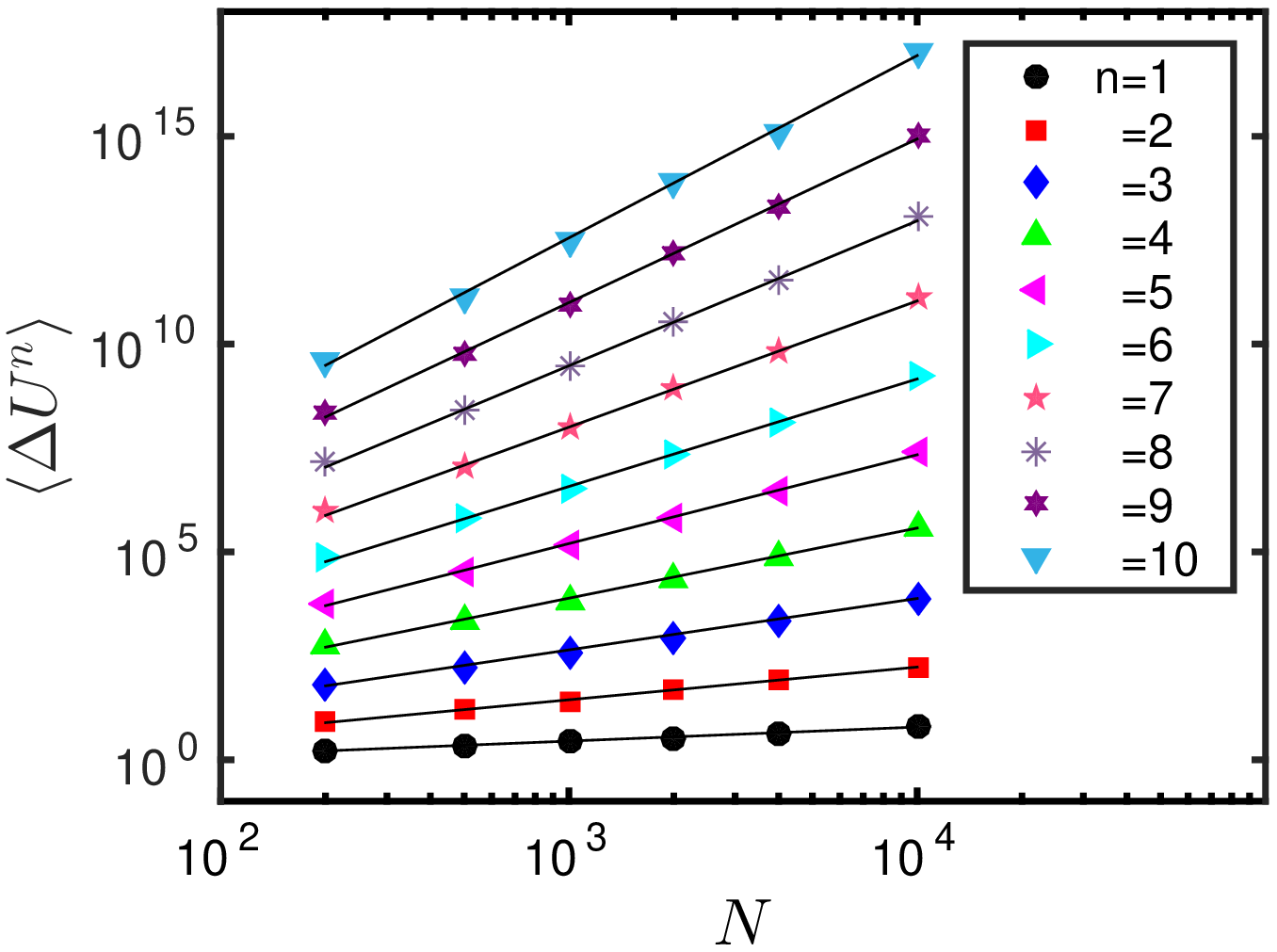}
\caption{Scaling plots for the moments defined in Eq.~\ref{posscale}.}
\label{positive}
\end{figure}

To begin the discussion of the statistics of stress and energy drops we summarize briefly what is known.
Most pertinent are the scaling laws for the mean stress and energy drops as a function of the system size,
\begin{equation}
\langle \Delta \sigma \rangle \sim N^\beta \ , \langle \Delta U \rangle \sim N^\alpha \ .
\label{posscale}
\end{equation}
It was argued \cite{10KLP} that exact scaling relations imply exact values for the scaling exponents, i.e. $\alpha=1/3$ and $\beta=-2/3$. Indeed, measurements
of the mean drops in the steady-state regime confirm these predictions to high accuracy, see Fig.~\ref{positive} and Table I.
This figure shows the system size dependence of not only the mean drop size, but also of the $n$'th moments of the drop sizes
as defined by
\begin{equation}
\langle (\Delta\sigma)^n\rangle\sim N^{\beta_n} \ , \quad \langle (\Delta U)^n\rangle\sim N^{\alpha_n}\ ,  n=1,2 \dots 10 \ .
\end{equation}
In Table I the measured values of the exponents $\alpha_n$ and $\beta_n$ are displayed. These are obtained from
straightforward least-squares fit to the data shown in Fig.~\ref{positive}.
\begin{small}
\begin{table}
\label{alnbetn}
\begin{center}
\begin{tabular}{ |c|c|c|c|c|c|c|c|c|c|c|}
\hline
 n &1 & 2 & 3 & 4 & 5 & 6 & 7 & 8 & 9 & 10 \\
 \hline
$\alpha_n$ &0.33 &0.77 & 1.22 & 1.68& 2.14&2.59&3.05&3.50&3.94&4.38 \\
 $\beta_n$&  -0.65 & -1.13  &-1.58  & -2.02 & -2.46&-2.90&-3.34&-3.78&-4.22&-4.66 \\
\hline
\end{tabular}
\end{center}
\caption{Scaling exponents $\alpha_n$ and $\beta_n$ for $n=1,2,\cdots, 10$.}
\end{table}
\end{small}

\begin{figure}
\includegraphics[width=7cm]{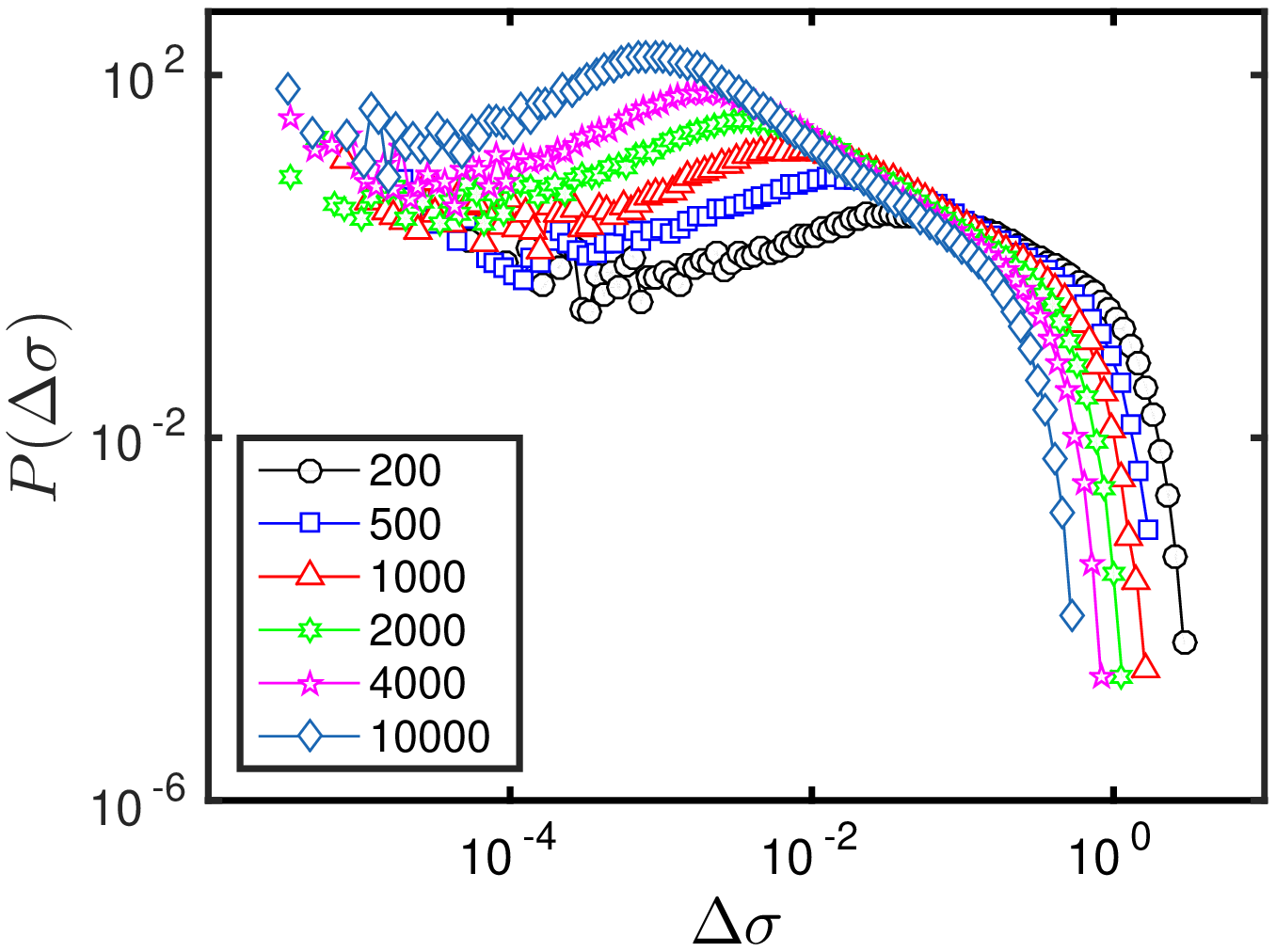}
\includegraphics[width=7cm]{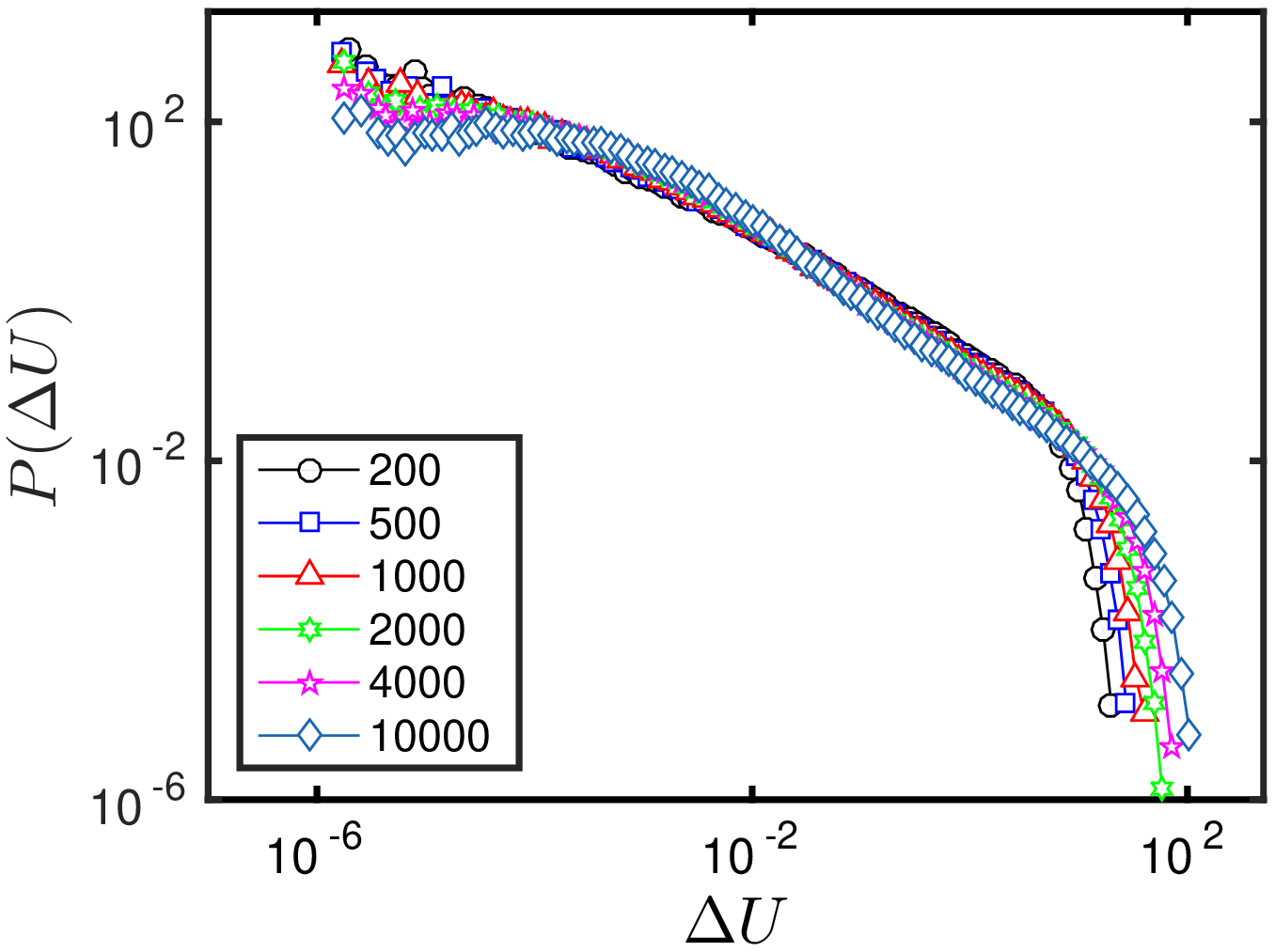}
\caption{The distribution functions $P(\Delta \sigma;N)$ and $P(\Delta U;N)$ of the stress (upper panel) and energy (lower panel) drops for
different system sizes}
\label{pdf}
\end{figure}

To understand the scaling exponents we need to examine the probability distribution functions (pdf's) of the stress and
energy drops. Collecting enough data allows us to present accurate pdf's which we denote as $P(\Delta \sigma;N)$ and
$P(\Delta U;N)$ respectively. These are displayed in Fig.~\ref{pdf}. In many studies of such pdf's one tends
to fit a power low to the apparent straight regimes shown in such plots. We will show here that this is not necessarily
a very useful procedure. It is much more rewarding to try to determine and to understand
\begin{figure}
\includegraphics[width=7cm]{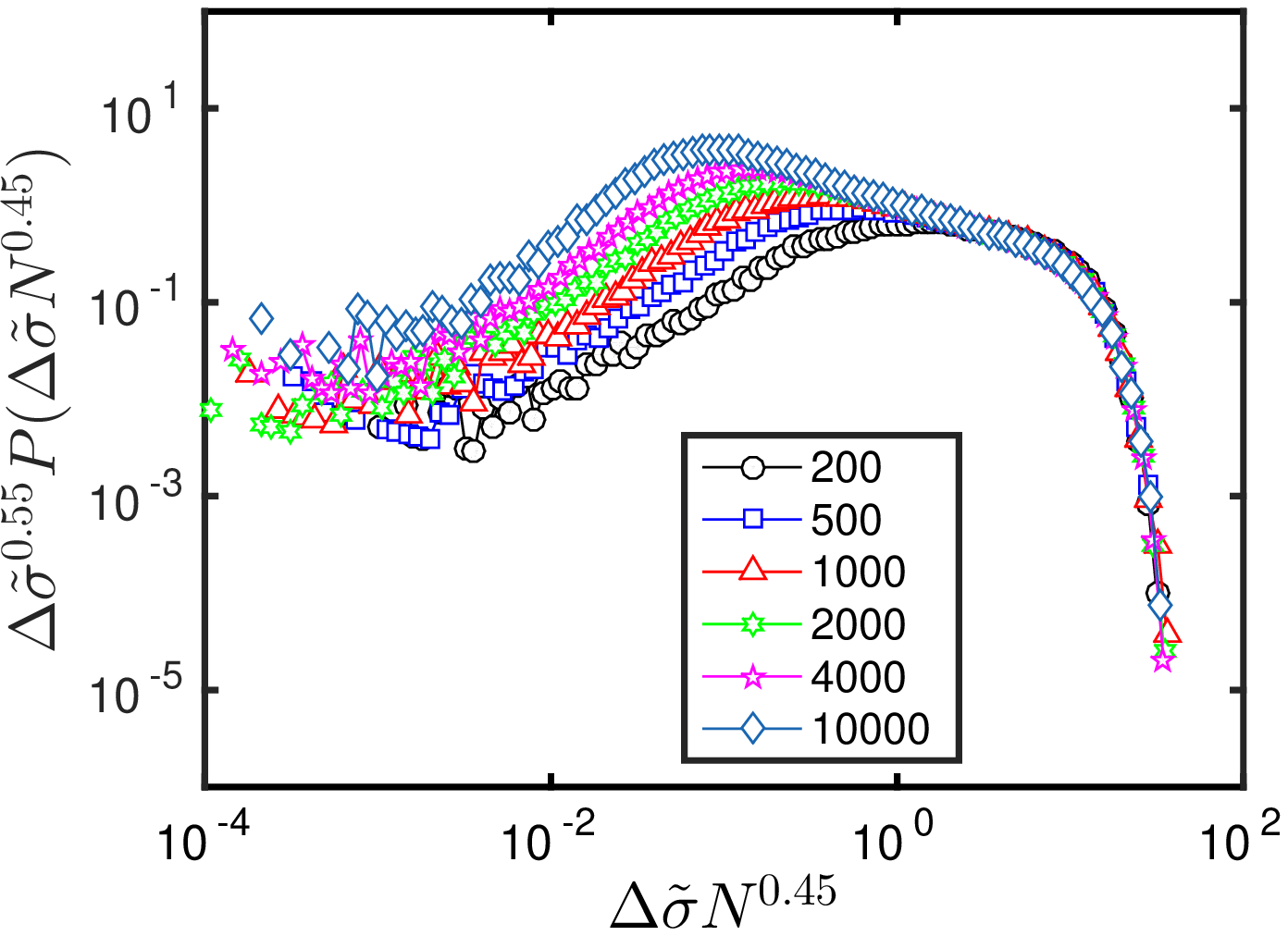}
\includegraphics[width=7cm]{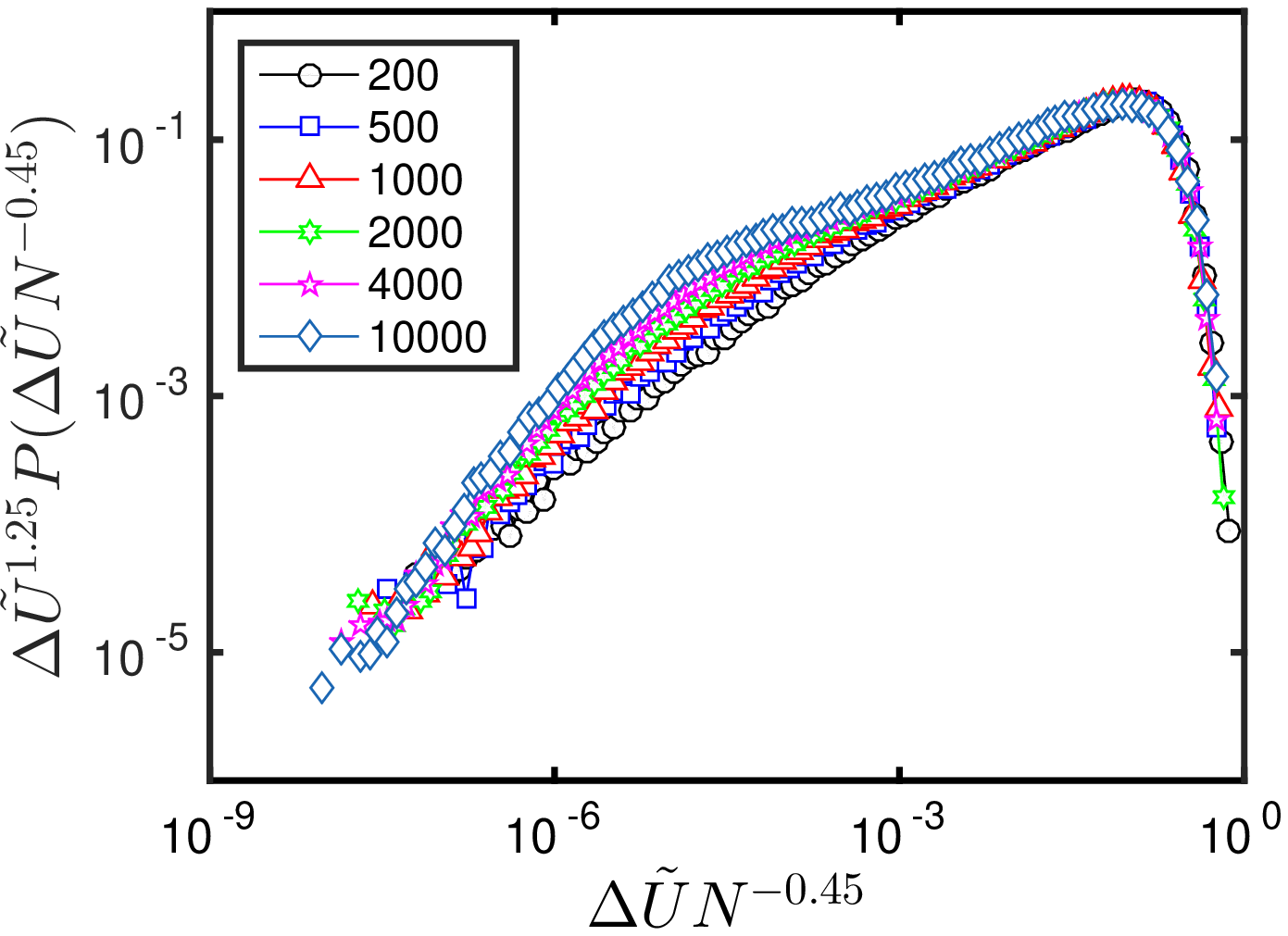}
\includegraphics[width=7cm]{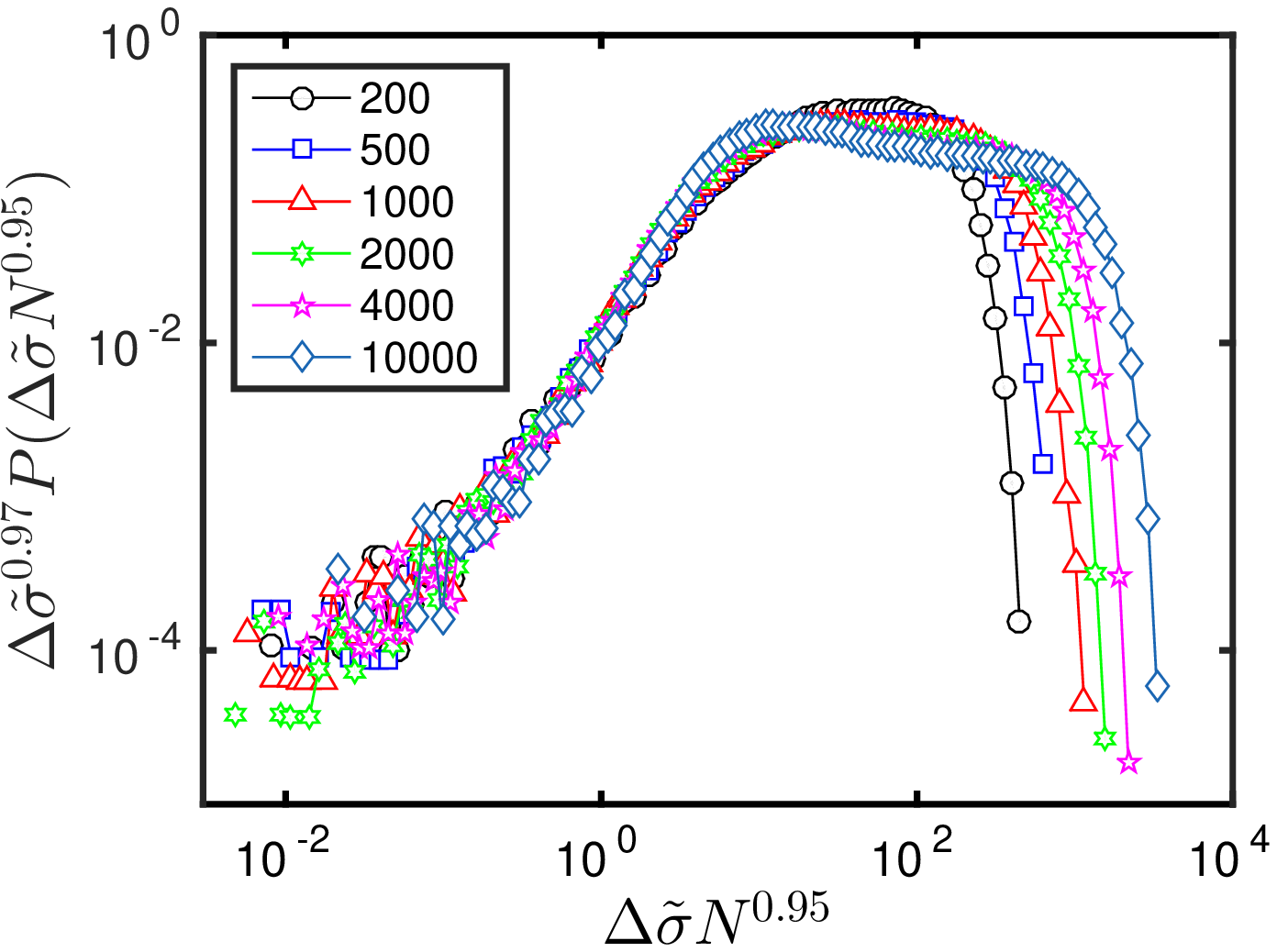}
\caption{Rescaled pdf's stressing data collapse for the {\it larger} drops of $\Delta \sigma$ and $\Delta U$ (upper and middle panel),
and stressing collapse for the {\it smaller} drops of $\Delta \sigma$ (lower panel). Note that the data collapse of one part is on the expense of the other.}
\label{fullscale}
\end{figure}
the scaling exponents by trying to rescale the pdf's and to collapse the data
for all the system sizes on one curve. It immediately turns out that it is not possible to rescale the entire pdf's
to accomplish such a data collapse: one needs to decide which part of the pdf to collapse. Large drops and
small drops display different scaling properties!

\begin{figure}
\includegraphics[width=7cm]{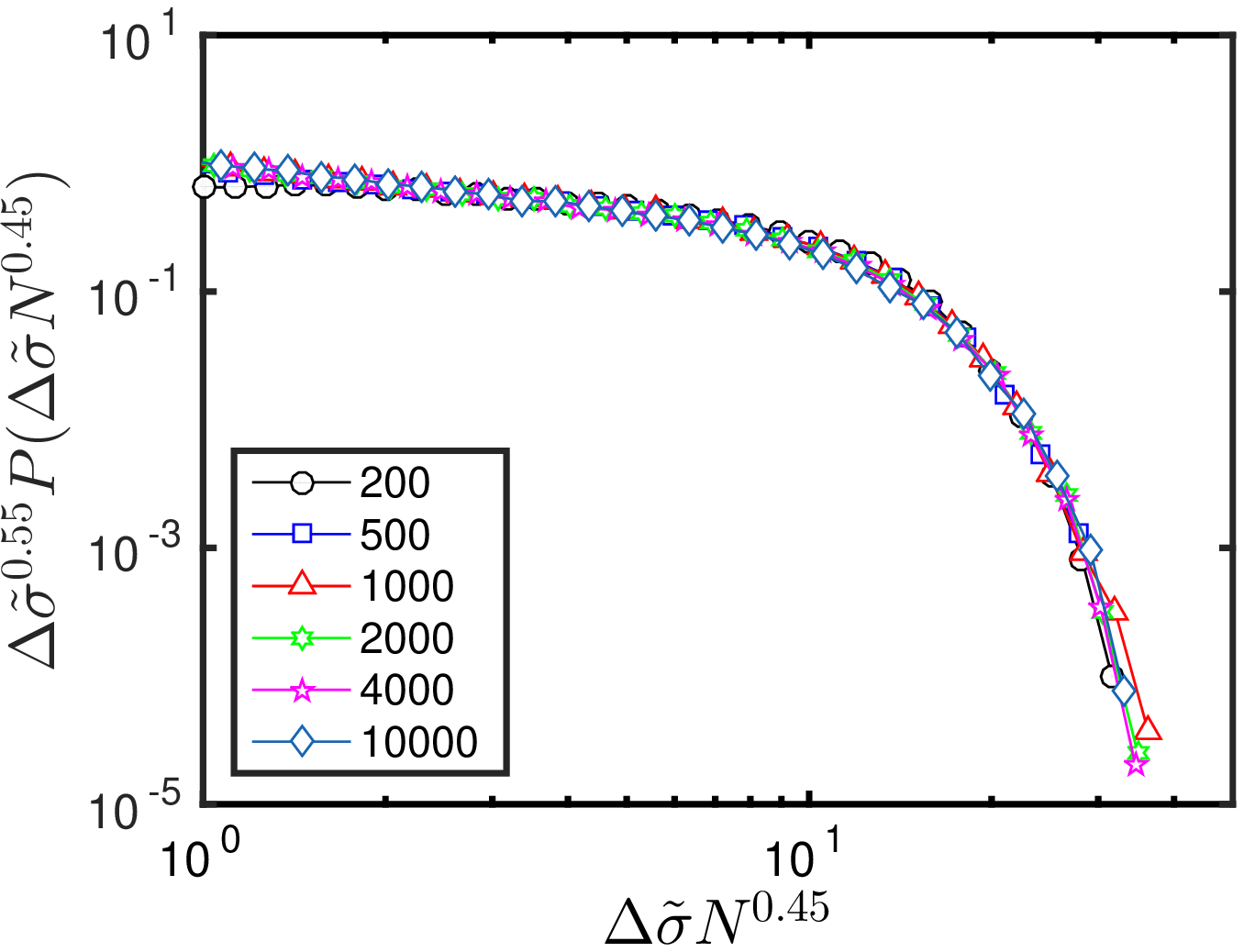}
\includegraphics[width=7cm]{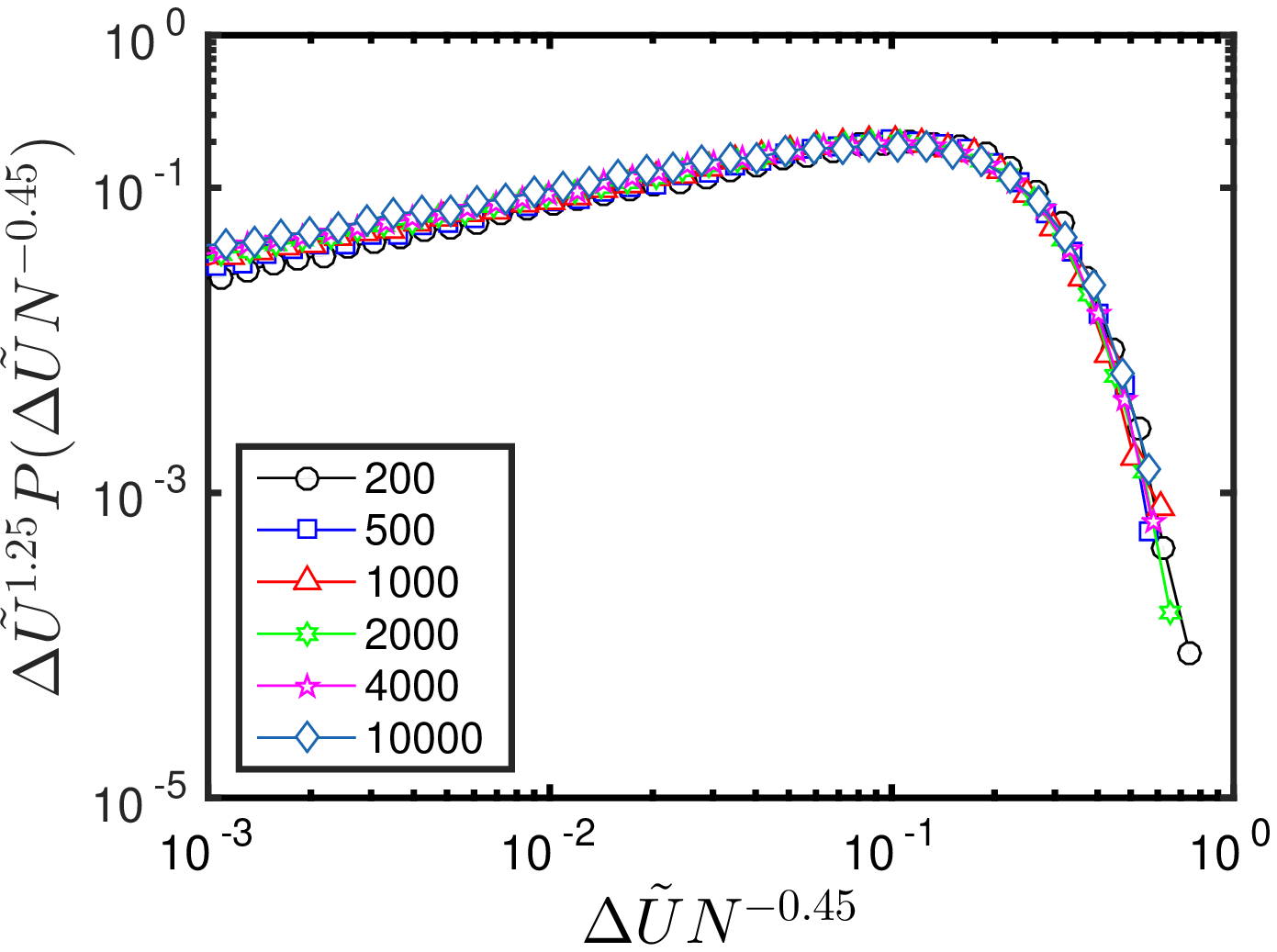}
\caption{Rescaled pdf's for the stress and energy drops with data collapse achieved in the range of
large stress and energy drops.}
\label{rescale-large}
\end{figure}

To make the point clear we present in Fig.~\ref{rescale-large} the results of the following rescaling. First
we rescale $\Delta \sigma$ and $\Delta U$ by the mean values in the steady state stress $\sigma_\infty$ and energy per particle $U_\infty$:
\begin{equation}
\Delta \tilde\sigma \equiv \Delta\sigma/\sigma_\infty \ ,\quad \Delta \tilde U \equiv \Delta U/|U_\infty| \ .
\end{equation}
Next we represent the pdf using a scaling function of dimensionless variables:
\begin{eqnarray}
P(\Delta \tilde\sigma)& = &(\Delta \tilde\sigma)^{\zeta_\sigma} f_\sigma (\Delta \tilde\sigma N^{\theta_\sigma}) \nonumber\\
P(\Delta \tilde U) &=& (\Delta \tilde U)^{\zeta_U} f_U (\Delta \tilde U N^{-\theta_U}) \ .
\label{defexp}
\end{eqnarray}
In trying to optimize the data collapse by selecting appropriate values for the scaling exponents $\zeta_\sigma, \theta_\sigma, \zeta_U$ and $\theta_U$, we discover that we can collapse either the part of the pdf that pertains to large or to
small drops. By choosing $\zeta_\sigma=0.55, \theta_\sigma=0.45, \zeta_U=1.25$ and $\theta_U=0.45$
we can achieve an apparent good collapse of the pdf's for {\em larger} values of the stress and energy drops, on the expense of a divergence of the rescaled pdf's from each other for small values of the drops (see Fig.~\ref{fullscale}).
As can be gleaned from Fig.~\ref{fullscale}, this divergence is most apparent in the
stress drop distribution, and less so in the energy drop pdf. The reason and implications of this difference will be
discussed below.
\subsection{Scaling exponents of the positive moments: numerical results}
Since the scaling exponents of the positive moments of the stress and energy drops are expected to be dominated
by large rather than small values of the drops, we can use now the rescaled versions of the pdf's to predict the
values of the scaling exponents shown in Table I. Using the forms Eqs.~\ref{defexp} we compute the predictions
\begin{equation}
\beta_n=\theta_\sigma(-n+\zeta_\sigma-1) \ , \alpha_n=\theta_U(n-\zeta_U+1) \ .
\label{predictions}
\end{equation}
The reader can verify that these prediction are in apparent agreement with the measured values reported in
table I up to errors of the order of $0.02$ in $\alpha_n$ and up to $0.04$ in $\beta_n$. The conclusion is
that the scaling behavior of the positive moments is ``simple scaling" resulting in a linear law
for $\alpha_n$ and $\beta_n$ which is determined by two independent numbers for each set of moments. These
four independent numbers seem to be non-trivial. We will show below that this is a result of the direct
numerical analysis but in fact a simpler picture is going to emerge once the physics is considered.
\subsection{Small drops and negative moments}

\begin{figure}
\includegraphics[width=7cm]{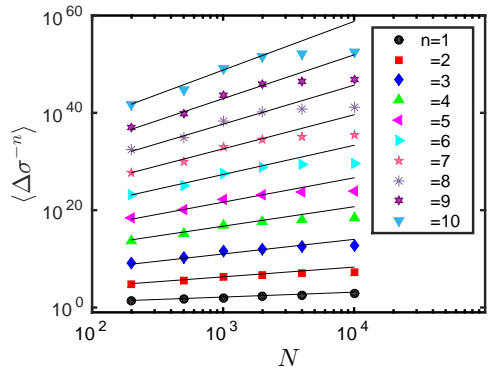}
\caption{The system size dependence of the negative moments of the stress drops according
to Eq.~\ref{defeps}.}
\label{negmom}
\end{figure}

\begin{figure}
\includegraphics[width=7cm]{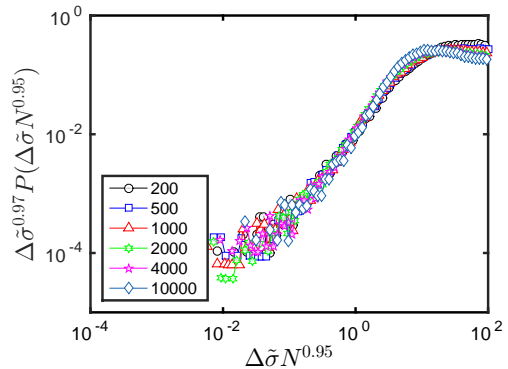}
\caption{The numerical choice of the rescaled pdf of stress drops with attention to the smaller drops.}
\label{smallsig}
\end{figure}

It is interesting and relevant that the pdf restricted to the smaller stress drops does not conform
to the scaling data collapse proposed for the large drops. To study this further we address now the {\em negative}
moments of stress drops, i.e.
\begin{equation}
\langle (\Delta\sigma)^{-n}\rangle \sim N^{\epsilon_n} \ .
\label{defeps}
\end{equation}
The data for the system size dependence of these negative moments is shown in Fig.~\ref{negmom}.
From the best linear fits to the log-log plots of the moments vs the system size we extract the
set of exponents presented in Table II. One should note that the accuracy of the scaling laws
deteriorates for the higher order negative moments and the determination of the scaling law
for $n>6$ becomes less definite. Nevertheless we proceed to rescale the pdf of the stress drops, restriced
to the small drops
\begin{small}
\begin{table}
\label{epsn}
\begin{center}
\begin{tabular}{ |c|c|c|c|c|c|c|c|c|c|c|}
\hline
 n &1 & 2 & 3 & 4 & 5 & 6 & 7 & 8 & 9 & 10 \\
 \hline
$\epsilon_n$ &0.94 &1.78 & 2.65 & 3.52& 4.38&5.25&6.12&6.98&7.85&8.71 \\
\hline
\end{tabular}
\end{center}
\caption{Scaling exponents $\epsilon_n$ for $n=1,2,\cdots, 10$.}
\end{table}
\end{small}
to understand these exponents. The numerical fit for achieving data collapse is written as
\begin{equation}
P(\Delta \tilde\sigma) = (\Delta \tilde\sigma)^{0.97} f_\sigma (\Delta \tilde\sigma N^{0.95}) \ .
\label{scalesmall}
\end{equation}
The resulting data collapse is shown in Fig.~\ref{smallsig}. One should note that with the present choice
of rescaling exponents the parts of the pdf's pertaining to the larger stress drops do not collapse on each other (see Fig.~\ref{fullscale}).
Using the rescaled form of the pdf we can now compute the expected exponents $\epsilon_n$.
The resulting exponents are in reasonable agreement with the list shown in Table II, with errors
not exceeding 8-9\%. Below we will argue that the real picture is in fact much simpler.

The real nature of the small drops becomes apparent when we consider the small energy drops.
We consider now the pdf of the energy drops restricted to small values.
\begin{figure}
\includegraphics[width=7cm]{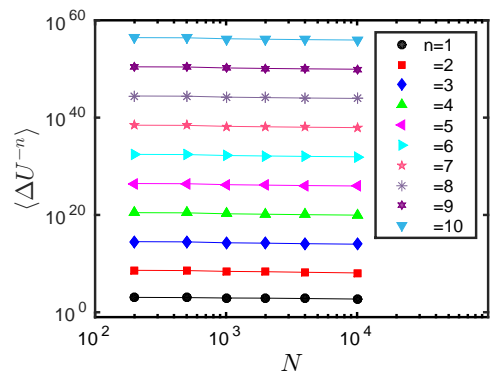}
\caption{The negative moments $\langle \Delta U^{-n}\rangle$ vs the system size. The lack of change in the slopes
indicate the the small energy drops are system size independent in agreement with a localized event.}
\label{smallu}
\end{figure}
As before, we can plot the negative moments $\langle \Delta U^{-n}\rangle$ as a function of the system size, shown in
Fig.~\ref{smallu}. Evidently there is no difference in the scaling exponents, meaning that the small energy
drops are system size independent, i.e. $\Delta U\sim N^0$. This is a clear indication that the small drops
are associated with localized events, forcing us at this point to consider the physical significance of the
analysis discussed so far.
\section{Physical significance and theoretical results}
\begin{figure}
\includegraphics[width=7cm]{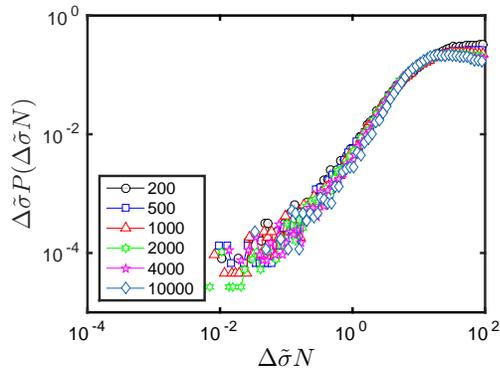}
\caption{The theoretically proposed rescaled pdf of stress drops with attention to the smaller drops.}
\label{corrsmall}
\end{figure}

Firstly, since the small energy drops are system size independent, we must associate the small drops
with localized events \cite{10KLP}. But then the stress drops, being intensive rather than extensive like the energy
drops, are expected to scale like $1/N$ if the energy drops are proportional to $N^0$. We then realize
that Eq.~(\ref{smallsig}) was only approximate, and that the exact representation of the pdf of the
small stress drops must read
\begin{equation}
P(\Delta \tilde\sigma) = (\Delta \tilde\sigma) f_\sigma (\Delta \tilde\sigma N) \ .
\label{corrscalesmall}
\end{equation}
Having realized this, we now check whether using the trivial scaling form (\ref{corrscalesmall}) collapses
the data as well as Eq.~(\ref{scalesmall}). The answer is yes, as can be inferred from Fig.~\ref{corrsmall}.
The quality of the data collapse is as good, if not better. We therefore can conclude that the small drops
occurring in the steady state are in fact regular plastic drops of the type that are prevalent
before yield. These are typical Eshelby quadrupolar events which indeed are known to be associated with
and energy drop which is system-size independent. With this realization we can now correct the list
of exponents $\epsilon_n$, these should be $\epsilon_n=n$, very trivial, and the discrepancy with the list
in table II is simply due to numerical inaccuracies.

\begin{figure}
\includegraphics[width=7cm]{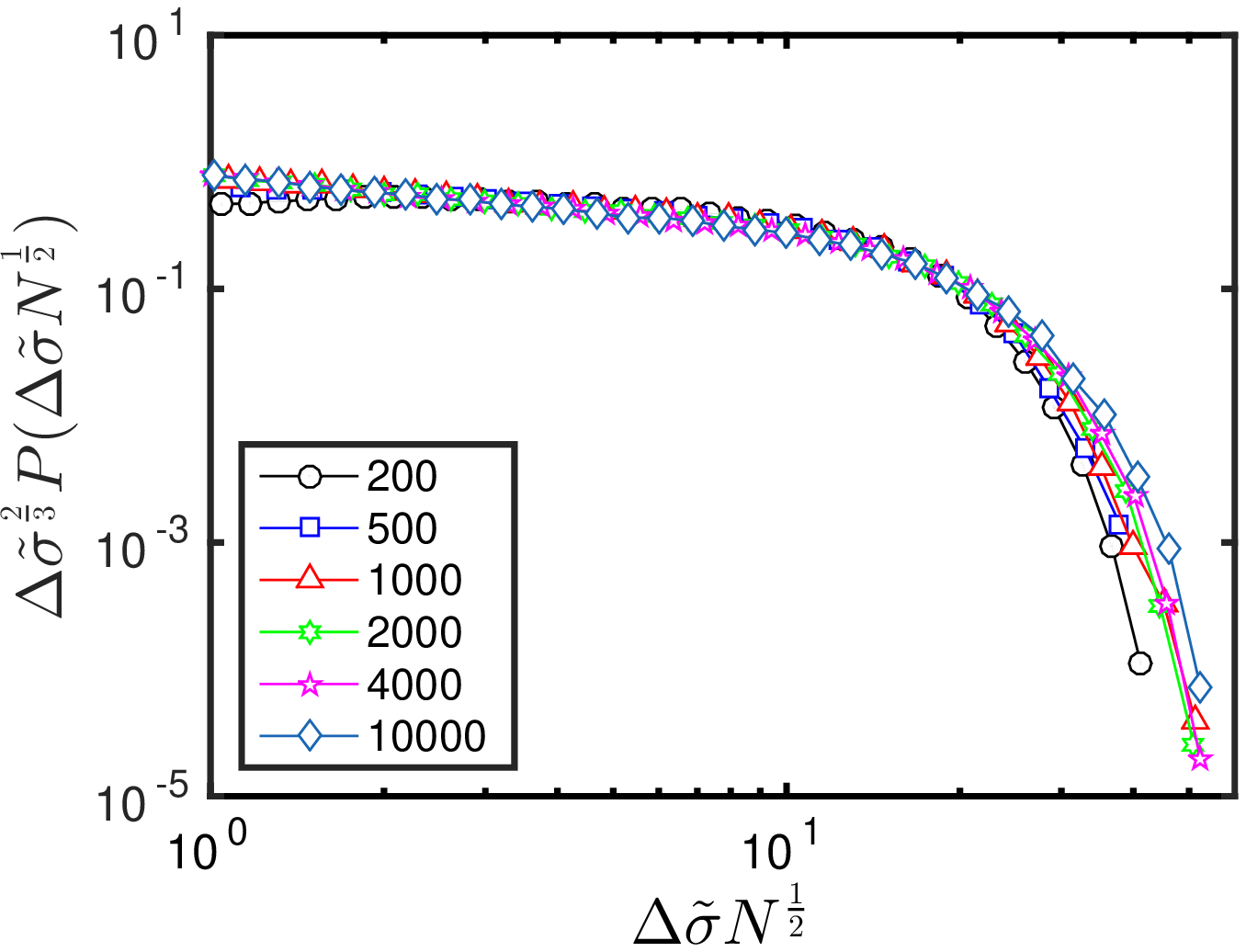}
\includegraphics[width=7cm]{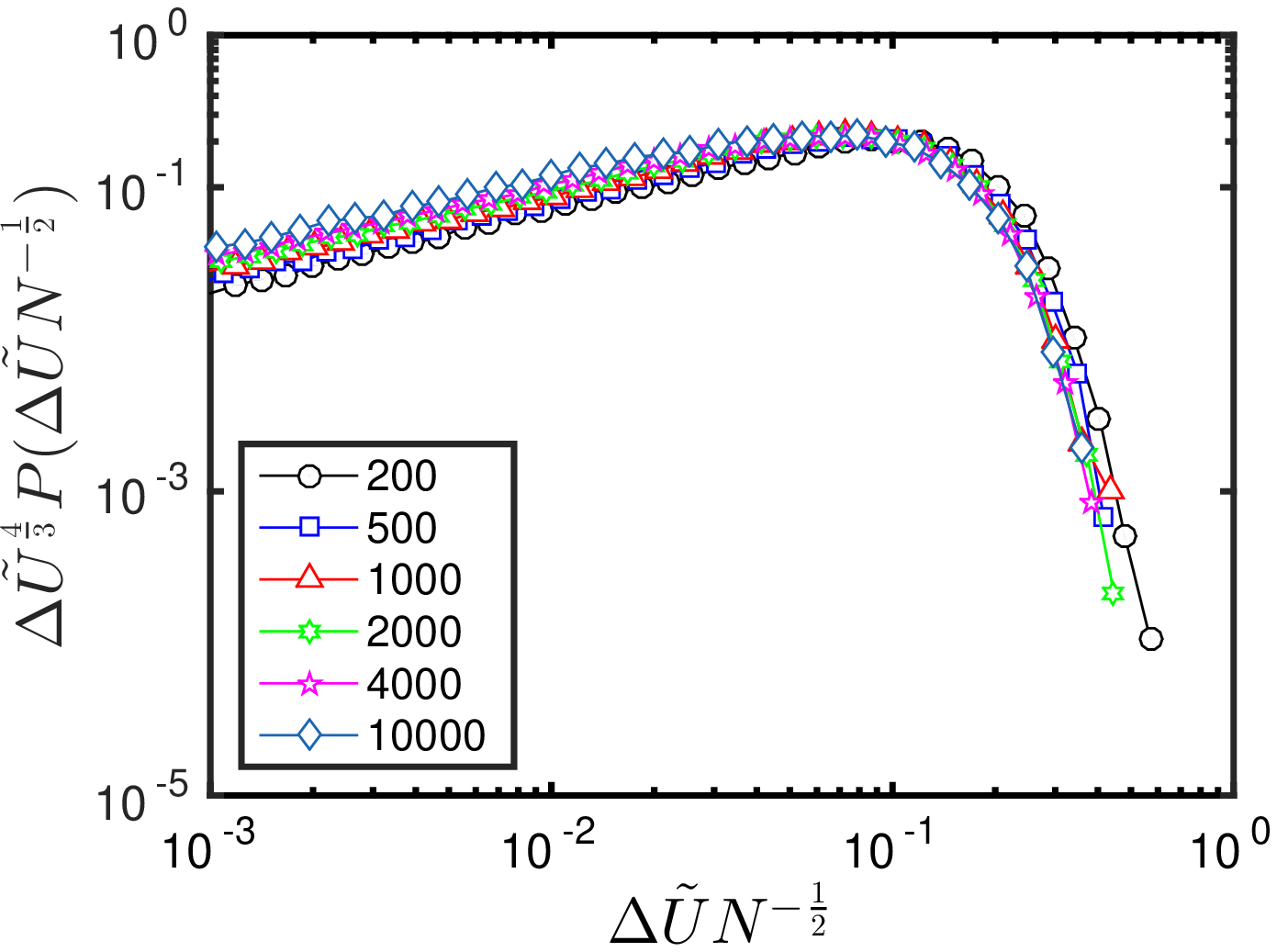}
\caption{Theoretical rescaled pdf's with attention to the large values of stress and energy drops.}
\label{final}
\end{figure}

This leaves us now with the large drops. But these must be the ``micro shear bands" that were identified and
discussed in some depth in Refs.~\cite{12DHP,13DGMPZ,13DHP}. In short, these are events that appear only after yield,
never before, and they are represented as a concatenated series of Eshelby quadrupoles that span the system.
These are in fact plastic events that organize the displacement field to sharply concentrate the shear
over a narrow band which traverses throughout the system. Accordingly the number of particles involved must
scale like $\sqrt{N}$. Accepting this, we must conclude that $\theta_U=1/2$. To get agreement with  $\alpha_1=1/3$
we must accordingly assign the theoretical value $\zeta_U=4/3$. Since the associate large stress drops are the
intensive counterparts of the large energy drops, we are led to the assignment $\theta_\sigma=1/2$ and to get
the exact $\beta_1=-2/3$ we end up with $\zeta_\sigma=2/3$. In Fig.~\ref{final} we show the final, theoretically
propose rescaled pdfs for the large stress and energy drops, and we conclude that they indeed collapse the data,
possibly not as well as the numerically optimized ones, but this discrepancy should be only taken as a warning
against straight numerics.
\begin{figure}
\includegraphics[width=7cm]{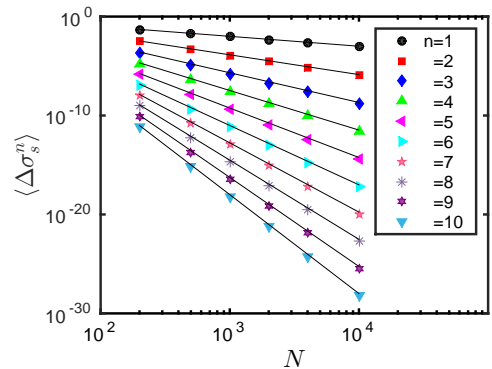}
\caption{The system size dependence of the positive moments of the small stress drops $\Delta \sigma_s\equiv\Delta \tilde \sigma N \leq 20$.}
\label{smallsigpos}
\end{figure}

To increase the confidence in the theoretical scaling predictions we can now analyze {\em separately} small or large
drops, and find the the moments of their fluctuations. Thus for example we can focus on stress drops that obey
$\Delta \tilde \sigma N \leq 20$ (see Fig. \ref{corrsmall}), and plot their positive moments, see Fig.~\ref{smallsigpos}.
The slopes of these plots are $-n$ with errors of less than 1\%. This is of course a direct consequence of the
trivial scaling function Eq.~\ref{corrscalesmall}.
\section{Concluding remarks}
We presented a detailed analysis of a typical ``stick-slip" serrated signal of stress and energy
drops in the post yield elasto-plastic steady state of a strain-controlled typical amorphous solid.
Numerical data analysis of the rather complex looking signal yielded pdf's whose scaling properties
appeared to call for a definition of a number of non-trivial scaling exponents. Further scrutiny however
reveled that much of the apparent complexity was due to the mixture between two types of stress and
energy drops: small, or system size independent, and large, which were system size dependent. The inter-penetration
of the two types of drops confused the statistical properties of each independent type of events, leading
to a complex looking scaling theory which eventually turned out to be unnecessary. At the end the statistics
is very simple, mixing events where $\Delta U\sim N^0$ and $\Delta\sigma \sim 1/N$ with system spanning
shear bands with $\Delta U\sim N^{1/2}$ and $\Delta\sigma \sim N^{-1/2}$. It is very likely that other
serrated signals measured in other ``stick-slip" systems may hide rich mixtures of physical processes in much
the same way presented here. Similar care in the analysis of such data should therefore be exercised.

\acknowledgments
This work had been supported in part by an ``ideas" grant STANPAS of the ERC and by the Minerva
Foundation, Munich, Germany. We acknowledge with thanks fruitful discussions with Prabhat K. Jaiswal.


\end{document}